\documentclass[11pt]{article}
\usepackage{amssymb}
\usepackage{graphics}
\usepackage{epsfig}
\usepackage{a4wide}

\begin{document}
\newcommand{\eebbgg}{$e^+e^- \to \gamma\gamma b\bar b$ }
\newcommand{\eebbggg}{$e^+e^- \to \gamma\gamma b\bar b g$ }

\title{QCD corrections to the $\gamma\gamma b\bar b$ production at the ILC }
\author{ Guo Lei, Zhang Ren-You, Ma Wen-Gan, and Han Liang \\
{\small Department of Modern Physics, University of Science and Technology}  \\
{\small of China (USTC), Hefei, Anhui 230026, P.R.China} }

\date{}
\maketitle \vskip 15mm

\begin{abstract}
The $e^+e^- \to \gamma\gamma b\bar b$ is an irreducible background
process in measuring the $H^0 \to \gamma\gamma$ decay width, if
Higgs boson is produced in association with a $Z^0$-boson which
subsequently decays via $Z^0 \to b\bar b$ at the ILC. In this
paper we study the impact of the ${\cal O}(\alpha_s)$ QCD corrections
to the observables of the $e^+e^- \to \gamma\gamma b\bar b$ process in
the standard model. We investigate the dependence of the
leading-order and ${\cal O}(\alpha_s)$ QCD corrected cross
sections on colliding energy and the additional jet veto schemes.
We also present the results of the LO and ${\cal O}(\alpha_s)$ QCD
corrected distributions of the transverse momenta of final
particles, and the invariant masses of $b\bar b$- and
$\gamma\gamma$-pair.
\end{abstract}


\vfill
\eject
\baselineskip=0.32in

\renewcommand{\theequation}{\arabic{section}.\arabic{equation}}
\renewcommand{\thesection}{\Roman{section}.}
\newcommand{\nb}{\nonumber}

\newcommand{\Dir}{\kern -6.4pt\Big{/}}
\newcommand{\Dirin}{\kern -10.4pt\Big{/}\kern 4.4pt}
\newcommand{\DDir}{\kern -7.6pt\Big{/}}
\newcommand{\DGir}{\kern -6.0pt\Big{/}}

\makeatletter      
\@addtoreset{equation}{section}
\makeatother       

\section{Introduction}
\par
The Higgs mechanism is an essential part of the standard model (SM)
\cite{abb1,abb2}, which gives masses to the gauge bosons and
fermions. Until now the Higgs boson has not been directly detected
yet in experiment. The LEP collaborations have established the lower
bound of the SM Higgs mass as $114.4~GeV$ at the $95\%$ confidence
level (CL) \cite{2}. The Fermilab Tevatron experiments have excluded
the SM Higgs boson with mass between $156$ and $177~GeV$ at $95\%$
CL \cite{mh-Fermilab}. Recently, the ATLAS and CMS experiments at
the LHC have provided the upper limits of the SM Higgs mass as
$130~GeV$ and $127~GeV$ at $95\%$ CL respectively, and there are
several Higgs like events around the locations of $m_H \sim
126~GeV$(ATLAS) and $m_H \sim 124~GeV$ (CMS)
\cite{mh-Atlas}\cite{mh-CMS}. Further searching for Higgs boson and
studying the phenomenology concerning its properties are still the
important tasks for the present and upcoming high energy colliders.

\par
After the discovery of the Higgs boson, the main tasks will be the
precise measurements of its couplings with fermions and gauge bosons
and its decay width \cite{abb6}. The future International Linear
Collider (ILC) is an ideal machine for conducting efficiently and
precisely the measurements for the standard model (SM) Higgs
properties. The ILC is designed with $\sqrt{s}=200\sim 500~GeV$ and
${\cal L}=1000~fb^{-1}$ in the first phase of operation \cite{abb7}.
The measurements of the Higgs-strahlung Bjorken process $e^+e^- \to
H^0 Z^0$ provide precision access to the studies of triple
interactions between Higgs boson and gauge bosons ($Z^0Z^0H^0$ and
$\gamma Z^0H^0$) \cite{abb9,ZH}. As both the Higgs boson and
$Z^0$-boson are unstable particles, we can only detect their final
decay products. For the $Z^0$-boson, the main decay channel is $Z^0
\to b\bar b$, whose branching fraction is $15.12\%$ \cite{databook}.
The Higgs coupling studies at the ILC usually can be carried out by
means of (i) $e^+e^- \to H^0Z^0 \to H^0 l^+l^-~ (l = e, \mu)$
process \cite{Brient}, (ii) $e^+e^- \to H^0Z^0 \to H^0q \bar q$, and
(iii) via $WW$-fusion $e^+e^- \to H^0\nu \bar{\nu}$ \cite{Boos}. In
the SM and beyond, such as the two-Higgs-doublet model (THDM) and
the minimal supersymmetric standard model (MSSM), the precise ILC
data for the Yukawa Higgs boson processes $e^+e^- \to
H_{SM}^0~(H^0,A^0)b \bar b$ are significant for probing the small SM
Yukawa bottom coupling and determining the ratio of the vacuum
expectation values $\tan\beta$ \cite{eebbh}. The $H_{SM}^0~(H^0,A^0)
b \bar{b}$ production events can be selected by tagging both
(anti)bottom jets. As for the light SM Higgs boson, its main decay
is the $H^0 \to b\bar b$ mode with a branch fraction about $90\%$,
but this decay mode would be difficult to detect accurately. The
rare diphoton Higgs decay channel is of great importance, since a
precise measurement of its width can help us to understand the
nature of the Higgs boson and may possibly provide hints for new
physics beyond the SM. This requires not only the precise
measurement for the diphoton Higgs decay width, but also accurate
predictions for new physics signal and its background. Fortunately,
the ILC instrument would provide excellent facilities in energy and
geometric resolutions of the electromagnetic detectors to isolate
the narrow $\gamma\gamma$ signal from the huge $\gamma\gamma$
continuum background. Ref.\cite{abb10} provides the conclusion that
a precision of $10\%$ on the partial decay width of $H^0 \to \gamma
\gamma$ can be achieved at the ILC by the help of an excellent
calorimeter.

\par
The calculations for $e^+e^- \to \gamma \gamma f\bar f$ reaction at
the tree-level are given in Ref.\cite{Boos}, and the study for
measuring the branching ratio of $H^0 \to \gamma\gamma$ at a linear
$e^+e^-$ collider is provided in Ref.\cite{abb10}. There it is
demonstrated that the ability to distinguish Higgs boson signature
at linear $e^+e^-$ colliders, crucially depends on the understanding
of the signature and the corresponding background with
multi-particle final states. If we choose the $Z^0H^0$ production
events at the ILC with the subsequent $H^0\to\gamma\gamma$ and
$Z^0\to b\bar b$ decays, we obtain the events with $b\bar
b\gamma\gamma$ final state, and the $e^+e^- \to b\bar b\gamma\gamma$
process becomes an important irreducible background of $Z^0H^0$
production. Our calculation shows the integrated cross section for
the $e^+e^- \to b\bar b\gamma\gamma$ process can exceed $30~fb$ at
the $\sqrt{s}=300~GeV$ ILC, more than thirty thousand $b\bar b
\gamma \gamma$ events could be obtained in the first phase of
operation, and then the statistical error could be less than $1\%$.
Therefore, it is necessary to provide the accurate theoretical
predictions for the \eebbgg process in order to measure the diphoton
decay width of Higgs boson at the future ILC.

\par
In this paper, we calculate the full ${\cal O}(\alpha_s)$ QCD
corrections to the process \eebbgg. In the following section we present
the analytical calculations for the process at the leading-order (LO)
and ${\cal O}(\alpha^4\alpha_s)$ order. The numerical results and
discussions are given in section III. Section IV summarizes the
conclusions.

\vskip 5mm
\section{Calculations}
\par
In both the LO and QCD one-loop calculations for the process
\eebbgg, we adopted the t'Hooft-Feynman gauge, if not stated
otherwise. We use the FeynArts3.4 package \cite{abb11} to generate
Feynman diagrams and their corresponding amplitudes. The
reductions of the output amplitudes are implemented by using the
developed FormCalc-6.0 package \cite{formloop}.

\vskip 5mm  {\bf (1) LO cross section}
\par
\begin{figure*}
\begin{center}
\includegraphics*[124pt,400pt][460pt,663pt]{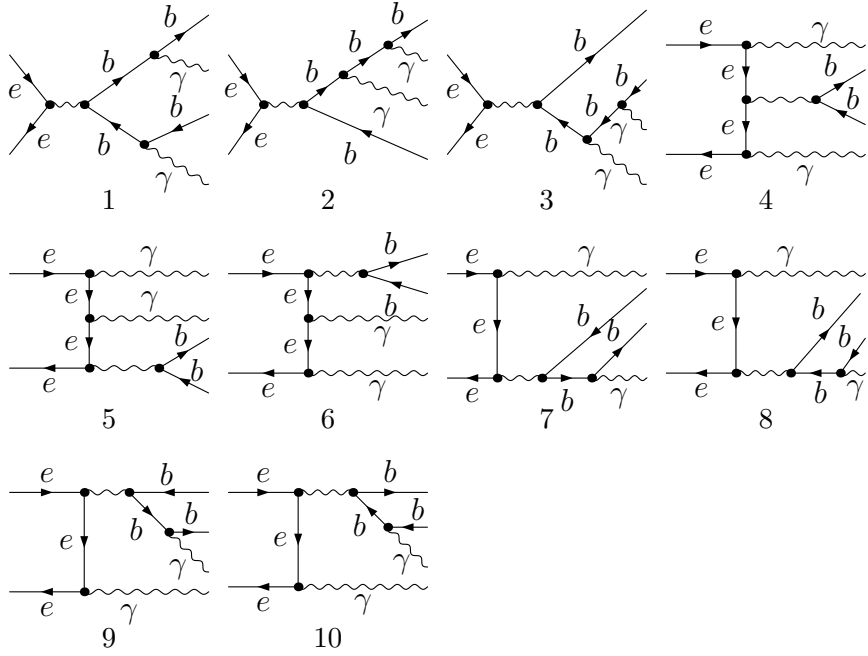}
\caption{\label{fig1} The generic Feynman diagrams at the LO for
the \eebbgg process. The internal wavy-line represents $\gamma$ or
$Z^0$-boson. The diagrams with exchanging the final two photons
are not drawn. }
\end{center}
\end{figure*}

The $b\bar{b}$-pair production associated with two photons via
electron-positron collision at the tree-level is a pure electroweak
process. We denote this process as $ e^+(p_1)+e^-(p_2) \to
\gamma(p_3)+ \gamma(p_4)+b(p_5)+\bar b(p_6)$, where $p_i~(i=1-6)$
label the four-momenta of incoming positron, electron and outgoing
final particles, respectively. Because the Yukawa coupling of
Higgs/Goldstone to fermions is proportional to the fermion mass,
we ignore the contributions of the Feynman diagrams which involve
the Yukawa couplings between any Higgs/Goldstone boson and electrons.
There are 40 generic tree-level diagrams for the process \eebbgg,
some of them are depicted in Fig.\ref{fig1}. The internal wavy-line
in Fig.\ref{fig1} represents $\gamma$- or $Z^0$-boson.

\par
The differential cross section for the process \eebbgg at the LO is
expressed as
\begin{equation}
d\sigma_{LO}= \frac{(2 \pi)^4 N_c}{2!~4~\sqrt{(p_1\cdot
p_2)^2-m_e^4}}\overline{\sum}\left|{\cal M}_{LO}\right|^2
{d\Phi_4}, \label{Sig}
\end{equation}
where $N_c=3$, factor $\frac{1}{2!}$ comes from the two final
identical photons, and $d\Phi_4$ is the four-body phase space
element given by
\begin{equation} \label{PSelement}
d\Phi_4  = \delta^ {(4)}\left (p_1+p_2-\sum_{i=3}^{6}p_i \right
)\prod_{i=3}^{6} \frac{d^3\vec{p}_i}{(2\pi)^3 2E_i}.
\end{equation}
The summation in Eq.(\ref{Sig}) is taken over the spins of final
particles, and the bar over the summation recalls averaging over
initial spin states. In the calculations, the internal $Z^0$-boson
is potentially resonant, and requires to introduce the finite width
in propagators. Therefore, we consider $Z^0$-boson mass, the related
$W^{\pm}$-boson mass and the cosine squared of Weinberg weak mixing
angle ($\theta_W$) consistently being complex quantities in order to
keep the gauge invariance \cite{CMS}. Their complex masses and
Weinberg weak mixing angle are define as
\begin{equation}\label{CM}
\mu_X^2=m_X^2-im_X\Gamma_X, ~~(X=W,Z),~~~~~
c_W^2=\frac{\mu_W^2}{\mu_Z^2},
\end{equation}
where $m_{W}$, $m_{Z}$ are conventional real masses and
$\Gamma_{W}$, $\Gamma_{Z}$ represent the corresponding total widths,
and the propagator poles are located at $\mu_X$ on the complex
$p^2$-plane. Since the $Z^0$- and $W^{\pm}$-boson propagators are
not involved in the loops for the ${\cal O}(\alpha_s)$ QCD
corrections, we shall not meet the calculations of N-point integrals
with complex internal mass. In our LO and QCD one-loop level
calculations for the \eebbgg process, we put cuts on the transverse
momenta of the produced photons and (anti)bottom-quarks
($p_{T,cut}^{(\gamma)}$, $p_{T,cut}^{(b)}$), final photon-photon
resolution ($\Delta R_{\gamma\gamma}^{cut}$), bottom-antibottom
resolution ($\Delta R_{b\bar{b}}^{cut}$) and final
(anti)bottom-photon resolution ($ \Delta R_{b(\bar b)\gamma}^{cut}$)
(The definition of $\Delta R$ will be declared in the following
section). Then the LO cross section for the \eebbgg process is
IR-finite.

\vskip 5mm {\bf (2) ${\cal O}(\alpha_s)$  QCD corrections}
\par
The full ${\cal O}(\alpha_s)$ QCD corrections to the \eebbgg process
can be divided into two parts: ${\cal O}(\alpha_s)$  QCD virtual and
real gluon emission corrections. The ${\cal O}(\alpha_s)$ QCD
virtual corrections include the contributions of the self-energy,
triangle, box, pentagon and counterterm diagrams. Since we take
non zero bottom-quark mass, the virtual QCD corrections do not
contain any collinear infrared (IR) singularity, and only the soft
IR singularities are involved in the virtual corrections.
We adopt dimensional regularization scheme with
$D=4-2\epsilon$ to extract both UV and IR divergences which
correspond to the pole located at $D = 4$ ($\epsilon = 0$) on the
complex $D$-plane, and manipulate the $\gamma_5$ matrix in
$D$-dimensions by employing a naive scheme presented in
Ref.\cite{gamma5}, which keeps an anticommuting $\gamma_5$ in all
dimensions. The wave function of the external (anti)bottom-quark
field and its mass are renormalized in the on-mass-shell
renormalization scheme.

\par
By introducing a suitable set of counterterms, the UV singularities
from one-loop diagrams can be canceled, and the total amplitude of
these one-loop Feynman diagrams is UV-finite. In the renormalization
procedure, we define the relevant renormalization constants of
bottom-quark wave functions and mass as
\begin{eqnarray}\label{RenConstant}
\psi_{b,0}^{L}=\left(1+\frac{1}{2}\delta
Z_{b(g)}^L\right)\psi_{b,0}^L,~~
\psi_{b,0}^{R}=\left(1+\frac{1}{2}\delta
Z_{b(g)}^R\right)\psi_{b,0}^R,~~m_{b,0}=m_b+\delta m_{b(g)}.
\end{eqnarray}

\par
With the on-mass-shell renormalization conditions we get the ${\cal
O}(\alpha_{s})$ renormalization constants as
\begin{eqnarray}
\label{counterterm}
 \delta m_{b(g)} &=& \frac{ m_b}{2}
        \widetilde{Re} \left ( \Sigma_{b(g)}^{L}(m_{b}^{2}) +
               \Sigma_{b(g)}^{R}(m_{b}^{2}) +
                2 \Sigma_{b(g)}^{S}(m_{b}^{2})  \right ), \nb \\
\delta Z_{b(g)}^{L} &=& - \widetilde{Re}\Sigma_{b(g)}^{L}(m_{b}^{2})
    - m_{b}^2 \frac{\partial}{\partial p^2}
        \widetilde{Re}\left [ \Sigma_{b(g)}^{L}(p^2) +
      \Sigma_{b(g)}^{R}(p^2) + 2\Sigma_{b(g)}^{S}(p^2) \right ]
      |_{p^2=m_{b}^2},  \nb \\
\delta Z_{b(g)}^{R} &=& -\widetilde{Re}\Sigma_{b(g)}^{R}(m_{b}^{2})
-
    m_{b}^2 \frac{\partial}{\partial p^2}
        \widetilde{Re}\left [ \Sigma_{b(g)}^{L}(p^2) +
                \Sigma_{b(g)}^{R}(p^2) +
                2 \Sigma_{b(g)}^{S}(p^2)  \right ]
                |_{p^2=m_{b}^2},
\end{eqnarray}
where $\widetilde{Re}$ takes the real part of the loop
integrals appearing in the self-energies only, and the
unrenormalized bottom-quark self-energies at ${\cal O}(\alpha_s)$
are expressed as
\begin{eqnarray}
\Sigma_{b(g)}^{L}(p^2)&=&\Sigma_{b(g)}^{R}(p^2)=\frac{g_s^2}{6
\pi^2}\left(-1+2B_0[p^2,0,m_b^2]+2B_1[p^2,0,m_b^2]\right), \nb \\
\Sigma_{b(g)}^{S}(p^2)&=&\frac{g_s^2}{3
\pi^2}\left(1-2B_0[p^2,0,m_b^2]\right).
\end{eqnarray}

\par
The IR divergences from the one-loop diagrams involving virtual
gluon can be canceled by adding the real gluon emission correction.
We denote the real gluon emission process as $e^+(p_1)+e^-(p_2) \to
\gamma(p_3)+ \gamma(p_4)+b(p_5)+\bar b(p_6)+g(p_7)$, where a real
gluon radiates from the internal or external (anti)bottom quark
line. We employ both the phase space slicing (PSS) method \cite{Harris}
and the dipole subtraction method \cite{dipole} for
gluon radiation to combine the real and virtual corrections in order
to make a cross check. In the PSS method the phase space of gluon emission
process is divided by introducing a soft gluon cutoff ($\delta_s=2~\Delta
E_7/\sqrt{s}$). That means the real gluon emission correction can be
written in the form as $\Delta\sigma^{real}_{QCD}=\Delta \sigma^{
soft}_{QCD}+\Delta \sigma^{hard}_{QCD}$. In this work we take the
non zero mass of bottom-quark and no collinear singularity exists in
the ${\cal O}(\alpha_s)$ QCD calculation. Therefore, we do not need to set the
collinear cut $\delta_c$ in adopting PSS method. Then the full
${\cal O}(\alpha_{s})$ QCD correction to the process \eebbgg is
finite and can be expressed as
\begin{equation}\label{cs}
\Delta \sigma_{QCD}  =  \Delta \sigma^{{\rm vir}}_{QCD} +
 \Delta \sigma^{{\rm real}}_{QCD}.
\end{equation}

\par
We use our modified FormCalc6.0 programs \cite{formloop} to simplify
analytically the one-loop amplitudes involving UV and IR singularities, and
extract the IR-singular terms from one-loop integrals in the amplitudes
by adopting the expressions for the IR singularities in
one-loop integrals \cite{Beenakk}. The numerical evaluations of the IR safe
N-point$ (N\leq5)$ scalar integrals are implemented by using the
expressions presented in Refs.\cite{OneTwoThree,Four,Five}. The
tensor loop integrals are expressed in scalar integrals via
Passarino-Veltman(PV) reductions \cite{Passarino}.

\vskip 5mm
\section{Numerical results and discussions }
\par
In this section we present the numerical results and discussions of
the LO and QCD corrected cross sections and the kinematical
distributions of the final particles for the \eebbgg process at the
ILC by using non zero bottom-quark and electron masses fixed at
$m_b=4.68~GeV$, $m_e = 0.511~MeV$. For the complex masses of
$W^{\pm}$- and $Z^0$-boson in Eq.(\ref{CM}), the real parts, $m_W$
and $m_Z$, are set to be the on-shell physical masses of $W^{\pm}$
and $Z^0$, i.e., $m_W = 80.399~GeV$ and $m_Z = 91.1876~GeV$. The
decay widths of $W^{\pm}$ and $Z^0$, which are the imaginary parts
of the complex masses, are taken to be $\Gamma_W = 2.085~GeV$ and
$\Gamma_Z = 2.495~GeV$, respectively \cite{databook}. The fine
structure constant is set to be $\alpha(m_Z^2)^{-1}=127.916$, and
the strong coupling constant at the $Z^0$-pole has the value of
$\alpha_s(m_Z^2)=0.1176$. The running strong coupling constant,
$\alpha_s(\mu^2)$, is evaluated at the three-loop level
($\overline{MS}$ scheme) with five active flavors \cite{databook}.
For the definitions of detectable hard photon and (anti)bottom quark
we require the constraints of $p_T^{(\gamma)} \geq
p_{T,cut}^{(\gamma)}$, $p_T^{(b)} \geq p_{T,cut}^{(b)}$ ($p_T^{(\bar b)}
\geq p_{T,cut}^{(b)}$), $\Delta R_{\gamma\gamma}
\geq \Delta R_{\gamma\gamma}^{cut}$, $\Delta R_{b\bar b} \geq \Delta
R_{b\bar b}^{cut}$ and $\Delta R_{b(\bar b)\gamma} \geq \Delta
R_{b(\bar b)\gamma}^{cut}$, where we apply the jet algorithm
presented in Ref.\cite{R} to the final photons and
(anti)bottom-jets. In the jet algorithm of Ref.\cite{R} $\Delta R$
is defined as $(\Delta R)^2 \equiv (\Delta \phi)^2+(\Delta \eta)^2$
with $\Delta \phi$ and $\Delta \eta$ denoting the separation between
the two particles in azimuthal angle and pseudorapidity
respectively. We set the QCD renormalization scale being
$\mu=\sqrt{s}/2$ in the numerical calculations if no other
statement. In further numerical evaluations, we take the cuts for
final particles having the values as $p_{T,cut}^{(\gamma)}=10~GeV$,
$p_{T,cut}^{(b)}=20~GeV$, $\Delta R_{\gamma\gamma}^{cut}=0.5$ and
$\Delta R_{b\bar b}^{cut}= \Delta R_{b(\bar b)\gamma}^{cut}=1$
unless otherwise stated. In the calculations, we use the 'inclusive'
and 'exclusive' selection schemes for the events including an
additional gluon-jet. In 'inclusive' scheme there is no restriction
to the gluon-jet, but in the 'exclusive' scheme the three-jet events
satisfy the conditions of $p_T^{(g)} > 20~GeV$ and $\Delta
R_{gb(\bar b)} > 1$ are excluded.

\par
\begin{figure}[htbp]
\begin{center}
\includegraphics[scale=0.53,bb=35 32 459 347]{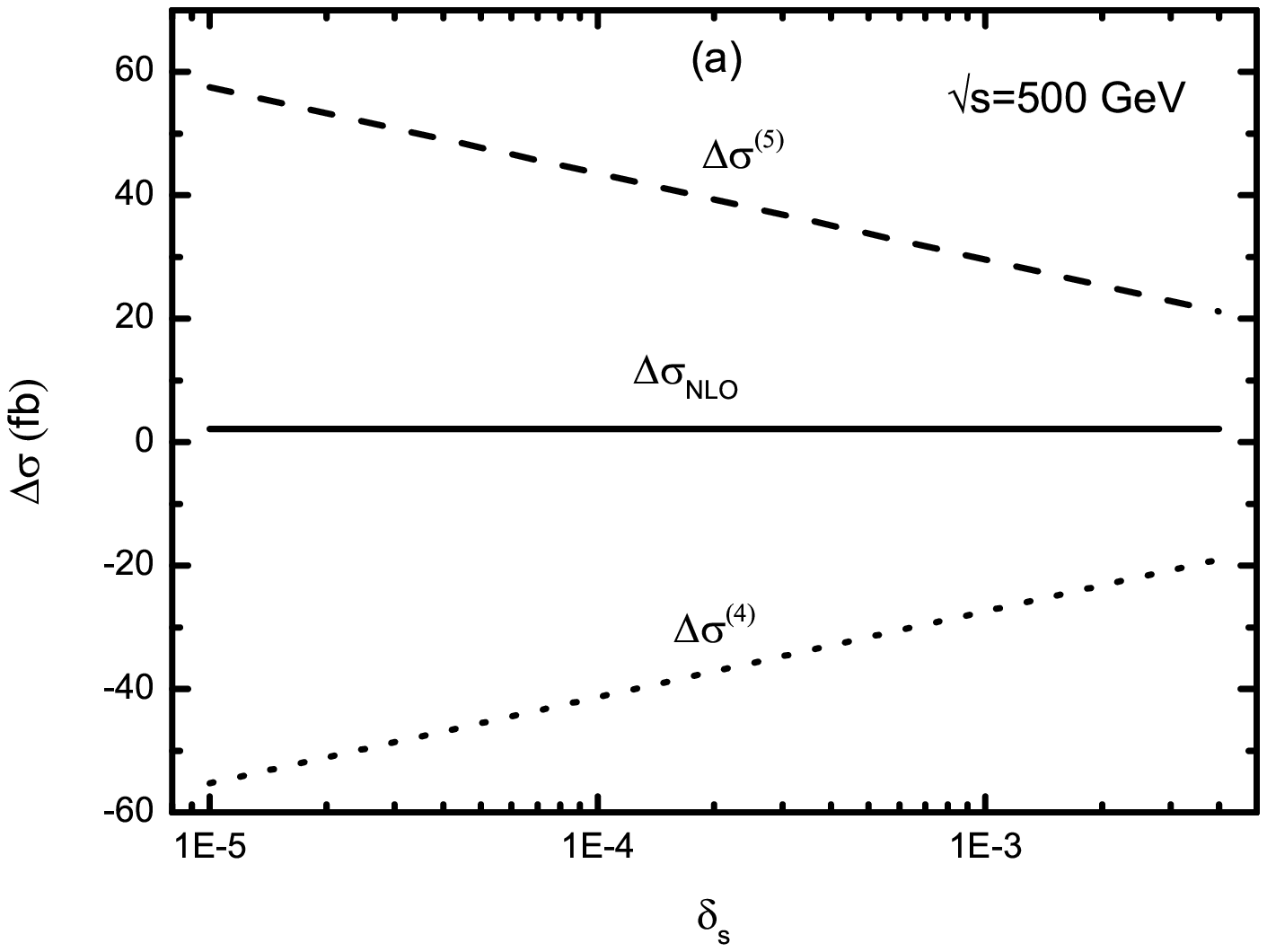}
\includegraphics[scale=0.53,bb=35 32 459 347]{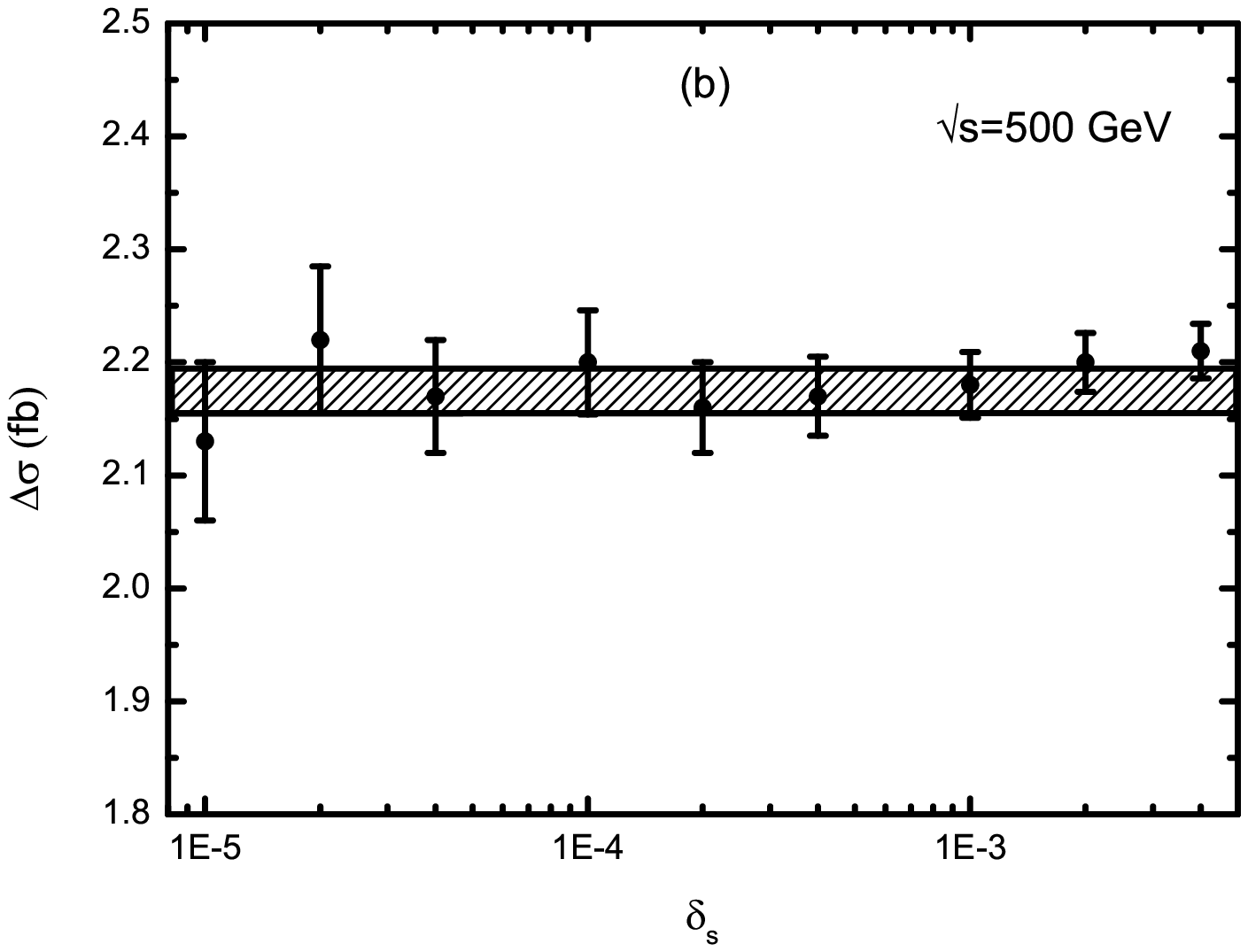}
\hspace{0in}%
\caption{\label{fig2} (a) The dependence of the correction components for
the \eebbgg process by adopting the phase space
slicing (PSS) method, $\Delta \sigma_{QCD}$, $\Delta \sigma^{(4)}$
and $\Delta \sigma^{(5)}$, on the soft cutoff $\delta_s$ at the ILC
by taking $\sqrt{s}=500~GeV$, $\mu=\sqrt{s}/2$ and the cut values mentioned above.
(b) The results for the full ${\cal O}(\alpha_s)$ QCD correction
$\Delta\sigma_{QCD}$ to the process \eebbgg by adopting the phase space
slicing (PSS) method together with Monte Carlo errors, and the shadowing region is for
the $\pm 1\sigma$ expected range of the results by adopting the dipole subtraction method. }
\end{center}
\end{figure}

We investigate the LO contribution from the $e^+e^- \to \gamma\gamma
Z^{0*} \to \gamma\gamma b \bar b$ channel as shown in
Figs.\ref{fig1}(4-6), and compare that part with the contribution
from all the diagrams for the \eebbgg process. We find that the
cross section for the $e^+e^- \to \gamma\gamma Z^{0*} \to
\gamma\gamma b \bar b$ channel is about $89\%-94\%$ of the LO total
cross section for the \eebbgg process, when the colliding energy
($\sqrt{s}$) goes from $200~GeV$ to $800~GeV$. It shows that the
dominant contributions are from the diagrams with resonant $Z^{0}$
exchanging, i.e., $e^+e^-\to \gamma\gamma Z^{0*}\to \gamma \gamma
b\bar b$ process, and the amplitude squared for \eebbgg process is
approximately proportional a Breit-Wigner function as $|M|^2 \propto
\frac{1}{(s_{56}-m_Z^2)^2 + m_Z^2\Gamma_z^2}$, where $s_{56}$ is the
squared invariant mass of $b\bar b$ pair. For this kind of
integration functions with large variation, an efficient and stable
Monte Carlo integration program is requested. We adopted our
in-house program to implement the four- and five-body phase space
integrations by applying the importance sampling for variable
$s_{56}$. In order to prevent numerical instability in tensor
integral reductions, we coded the numerical calculation programs in
Fortran77 with quadri-precision. With these programs the
precision and efficiency of Monte Carlo integration are greatly
improved. In order to verify the reliability of our numerical
results, we performed the following checks:

\begin{itemize}
\item The LO cross section for the process \eebbgg has
been calculated by adopting two independent packages and two gauges in the
conditions of $\sqrt{s}=500~GeV$ with the cuts of $p_{T,cut}^{(\gamma)}=10~GeV$
and $\Delta R_{\gamma\gamma}^{cut}=0.5$ for final photons,
and no cut for (anti)bottom quark. The numerical results are obtained as:
(1) By using CompHEP-4.5.1 program \cite{abb14}, we get
$\sigma_{LO}=29.05(4)~(fb)$ (in Feynman gauge) and
$\sigma_{LO}=29.02(3)~(fb)$ (in unitary gauge). (2) By using our
in-house $2 \to 4$ phase-space integration routine, we obtain
$\sigma_{LO}=29.03(3)~(fb)$ (in Feynman gauge) and
$\sigma_{LO}=29.06(3)~(fb)$ (in unitary gauge). We can see they are
all in good agreement within the statistic errors.

\item The independence of the full ${\cal
O}(\alpha_s)$ QCD correction on the soft cutoff $\delta_s$ is
confirmed numerically. Fig.\ref{fig2}(a) and Fig.\ref{fig2}(b)
demonstrate that the full ${\cal O}(\alpha_s)$ QCD correction to the
\eebbgg process at the ILC does not depend on the arbitrarily chosen
small value of the cutoff $\delta_s$ within the calculation errors,
where we take $\sqrt{s}=500~GeV$, $\mu=\sqrt{s}/2$,
$p_{T,cut}^{(\gamma)}=10~GeV$, $p_{T,cut}^{(b)}=20~GeV$, $\Delta
R_{\gamma\gamma}^{cut}=0.5$ and $\Delta R_{b\bar b}^{cut}= \Delta
R_{b(\bar b)\gamma}^{cut}=1$. In Fig.\ref{fig2}(a), the four-body correction
($\Delta\sigma^{(4)}$), five-body correction ($\Delta\sigma^{(5)}$)
and the full ${\cal O}(\alpha_s)$ QCD correction
($\Delta\sigma_{QCD}$) to the \eebbgg process are depicted as the
functions of the soft cutoff $\delta_s$ running from
$1\times 10^{-5}$ to $2\times 10^{-2}$. The amplified curve for the
full ${\cal O}(\alpha_s)$ correction is presented in
Fig.\ref{fig2}(b) together with calculation errors. The independence
of the total ${\cal O}(\alpha_s)$ QCD correction to the \eebbgg process
on the cutoff $\delta_s$ is a necessary condition that must be fulfilled
for the correctness of our calculations.

\item We adopt also the dipole subtraction method to deal with the IR
singularities for further verification. The results including
$\pm 1\sigma$ statistic errors are plotted as the shadowing region in
Fig.\ref{fig2}(b). It shows the results by using both the PSS method and
the dipole subtraction method are in good agreement. In further
numerical calculations we adopt the dipole subtract method.

\item The exact cancelations of UV and IR divergencies in our
${\cal O}(\alpha_s)$ QCD calculations are verified.

\end{itemize}

\par
\begin{figure}[htbp]
\begin{center}
\includegraphics[scale=0.5,bb=35 32 459 347]{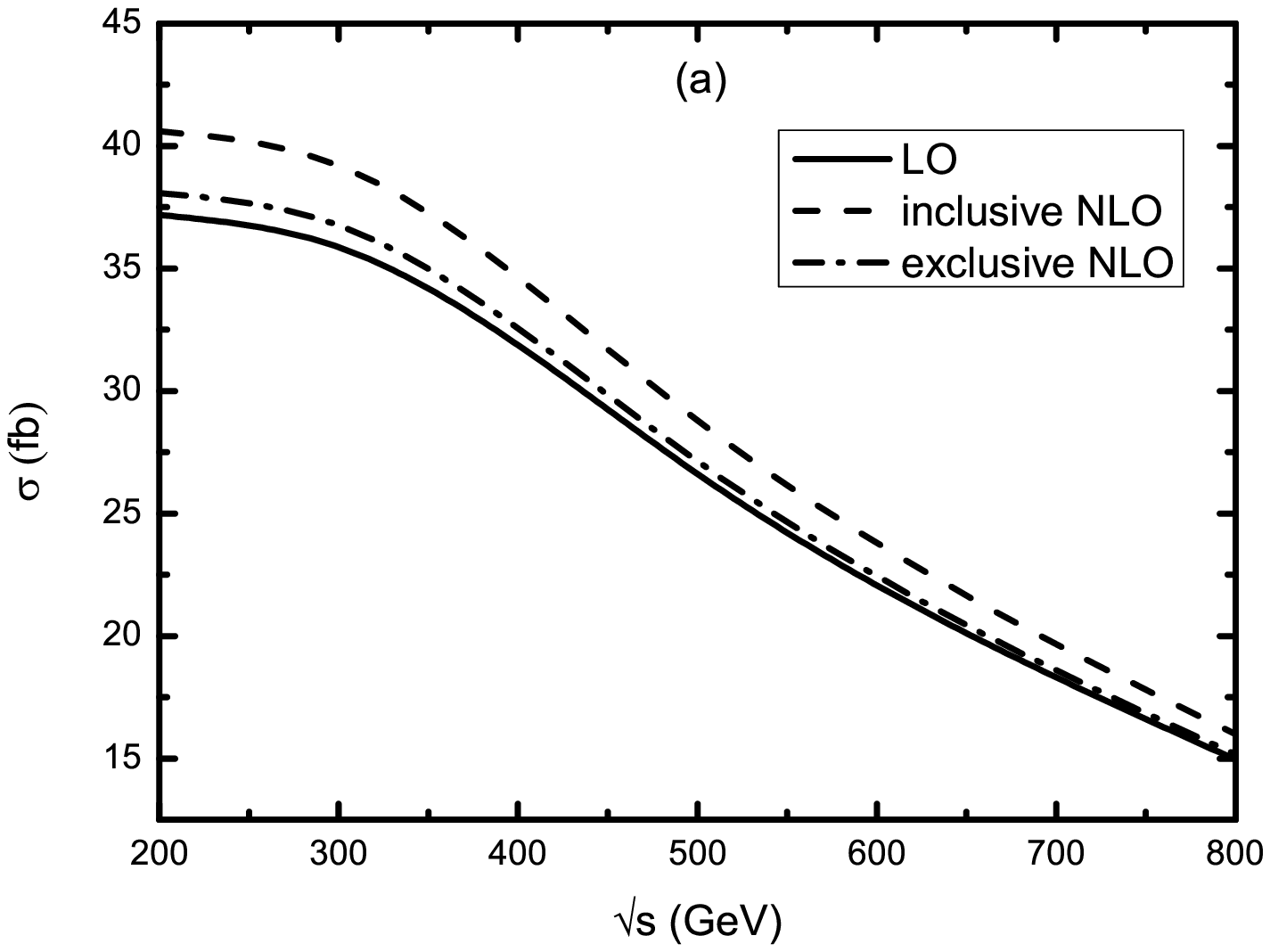}
\includegraphics[scale=0.5,bb=35 32 459 347]{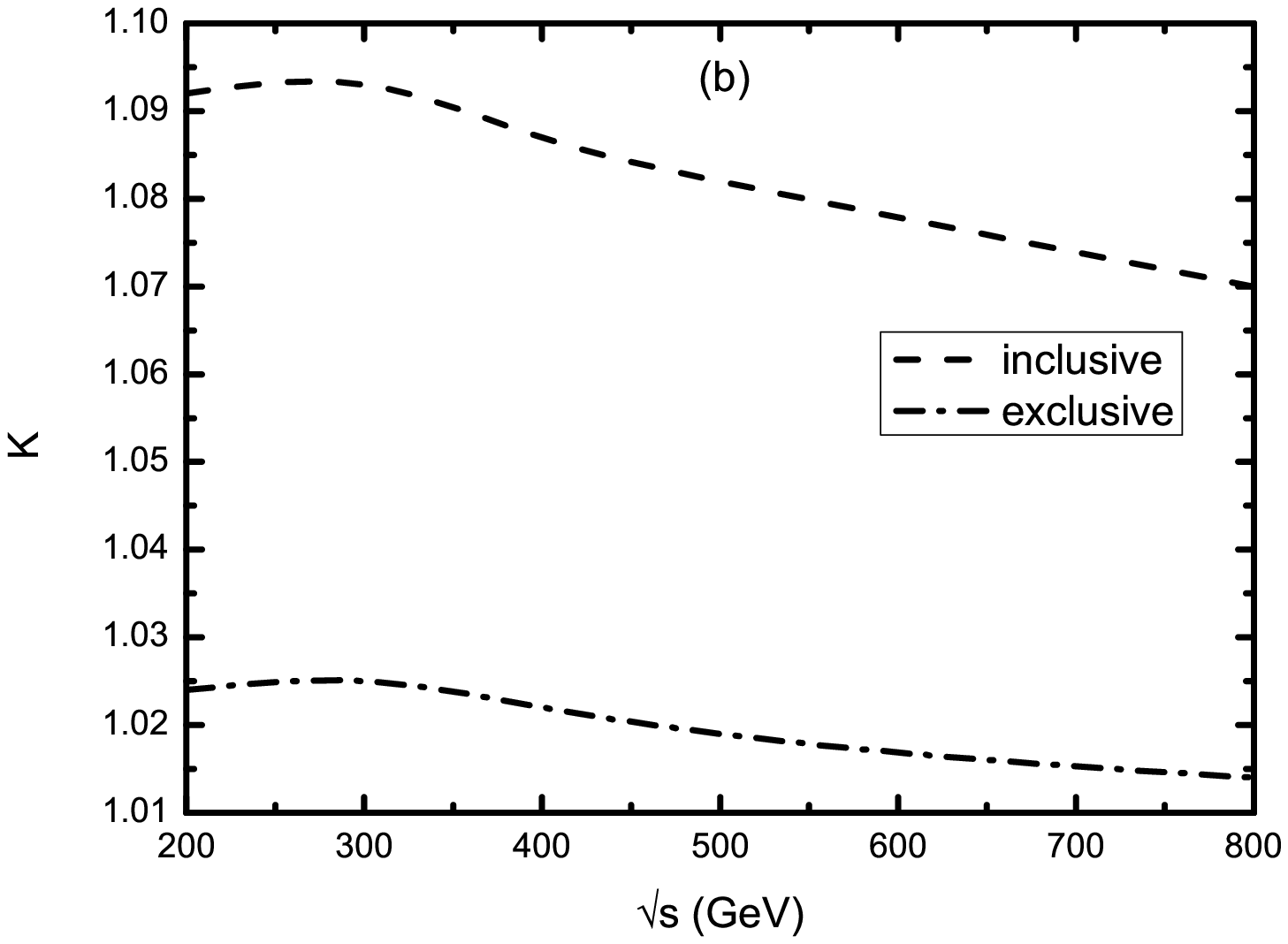}
\hspace{0in}%
\caption{\label{fig3} (a) The LO and ${\cal O}(\alpha_s)$ QCD
corrected cross sections with different event selection schemes for
the \eebbgg as the functions of the colliding energy $\sqrt{s}$
at the ILC with $\mu=\sqrt{s}/2$, $p_{T,cut}^{(\gamma)}=10~GeV$,
$p_{T,cut}^{(b)}=20~GeV$, $\Delta R_{\gamma\gamma}^{cut}=0.5$ and
$\Delta R_{b\bar b}^{cut}= \Delta R_{b(\bar b)\gamma}^{cut}=1$ for b-quarks
and photons. (b) The corresponding K-factors versus $\sqrt{s}$. }
\end{center}
\end{figure}

In Figs.\ref{fig3}(a,b) we depict the LO, ${\cal
O}(\alpha_s)$ QCD corrected cross sections and the corresponding
K-factors ($\equiv\sigma_{QCD}/\sigma_{LO}$) for the \eebbgg process
versus the colliding energy $\sqrt{s}$ at the ILC by taking
$\mu=\sqrt{s}/2$ and the cut set for b-quarks and photons mentioned above.
The figures show the QCD corrected results by adopting the 'inclusive'
and 'exclusive' three-jet event selection schemes, separately. We
list some of the data read out from these curves of
Figs.\ref{fig3}(a,b) in Table \ref{tab1}. We can see from the table
that the K-factor of the ${\cal O}(\alpha_s)$ QCD correction varies
quantitatively in the range of $1.092$ to $1.070$ for 'inclusive'
scheme, but in the range of $1.024$ to $1.014$ for 'exclusive'
scheme, when colliding energy $\sqrt{s}$ varies from $200~GeV$ to
$800~GeV$. As we know if the colliding energy is very large, the
dominant contribution for the \eebbgg process is from the
$\gamma\gamma Z^0$ production and followed by the real $Z^0$-boson
decay $Z^0 \to b \bar b$. Then the QCD K-factor for the \eebbgg
process is approximately equal to that for the later $Z^0$ boson
decay process. We make a comparison of the K-factors for the $e^+e^- \to
\gamma\gamma Z^0 \to \gamma\gamma b \bar{b}$ and the \eebbgg process
by using the 'inclusive' three-jet event selection scheme. We get the
K-factor of the $Z^0 \to b\bar b$ decay with the value of $1.069$,
and find it is agree with the K-factor of \eebbgg process at the ILC
with very high colliding energy, e.g., $K=1.070$ for $\sqrt{s}=800~GeV$.
From our calculations, we get the 'inclusive' ${\cal O}(\alpha_s)$ QCD
relative correction of $e^+e^-\to \gamma\gamma b\bar b$ at the $\sqrt{s}=300~GeV$ ILC is
about $9.3\%$, which is $2.4\%$ larger than the ${\cal O}(\alpha_s)$ QCD
correction estimated from the trivial ${\cal O}(\alpha_s)$ QCD corrections
for the decay $Z^0\to b\bar b$ convoluted with the production cross section for
$e^+e^-\to \gamma\gamma Z^0$. It shows that a complete ${\cal O}(\alpha_s)$ QCD
calculation for $e^+e^-\to \gamma\gamma b\bar b$ process is necessary, especially
in the first phase of ILC operation. We make a comparison for the renormalization
scale choices: i.e., $\mu=\sqrt{s}/2$ and $\mu=m_Z$. The former scale
value is close to $m_Z$ at the ILC running with a relative small
colliding energy. The ${\cal O}(\alpha_s)$ QCD corrections with the 'inclusive'
selection scheme at $\sqrt{s}=800~GeV$ are obtained as
$\sigma_{QCD}=16.01(2)~fb$, $K=1.071(3)$ for $\mu=m_Z$, and
$\sigma_{QCD}=15.99(2)~fb$, $K=1.070(3)$ for $\mu=\sqrt{s}/2$ as
shown in Table \ref{tab1}. It demonstrates that the ${\cal O}(\alpha_s)$ QCD
correction to the $e^+e^-\to b\bar b\gamma\gamma$ process is not
sensitive to these two renormalization scale choices.

\begin{table}
\begin{center}
\begin{tabular}{|c|c|c|c|c|c|c|}
\hline $\sqrt{s}$(GeV)  &200& 300 & 400 & 500 & 800  \\
\hline
$\sigma_{LO}(fb)$     & 37.19(1)    & 35.86(1)  & 31.86(1) & 26.59(1)   & 14.947(8) \\
$\sigma_{QCD}(fb)$(I) &  40.61(5)   & 39.20(5)  & 34.64(4) & 28.77(3)   & 15.99(2)  \\
$K$-$factor$(I)       &  1.092(3)   & 1.093(3)  & 1.087(3) & 1.082(3)   & 1.070(3)  \\
$\sigma_{QCD}(fb)$(II)&  38.08(5)   & 36.77(5)  & 32.57(4) & 27.10(3)   & 15.16(2)  \\
$K$-$factor$(II)      &  1.024(3)   & 1.025(3)  & 1.021(3) & 1.019(3)   & 1.014(3)  \\
\hline
\end{tabular}
\end{center}
\begin{center}
\begin{minipage}{15cm}
\caption{\label{tab1} The LO, ${\cal O}(\alpha_s)$ QCD corrected
cross sections and the corresponding K-factors with different jet veto
schemes at the ILC by taking $\mu=\sqrt{s}/2$, $p_{T,cut}^{(\gamma)}=10~GeV$,
$p_{T,cut}^{(b)}=20~GeV$, $\Delta R_{\gamma\gamma}^{cut}=0.5$
and $\Delta R_{b\bar b}^{cut}= \Delta R_{b(\bar b)\gamma}^{cut}=1$.
(I) For the 'inclusive' three-jet event selection scheme.
(II) For the 'exclusive' three-jet event selection
scheme.}
\end{minipage}
\end{center}
\end{table}

\par
\begin{figure}[htbp]
\begin{center}
\includegraphics[scale=0.53,bb=35 32 459 347]{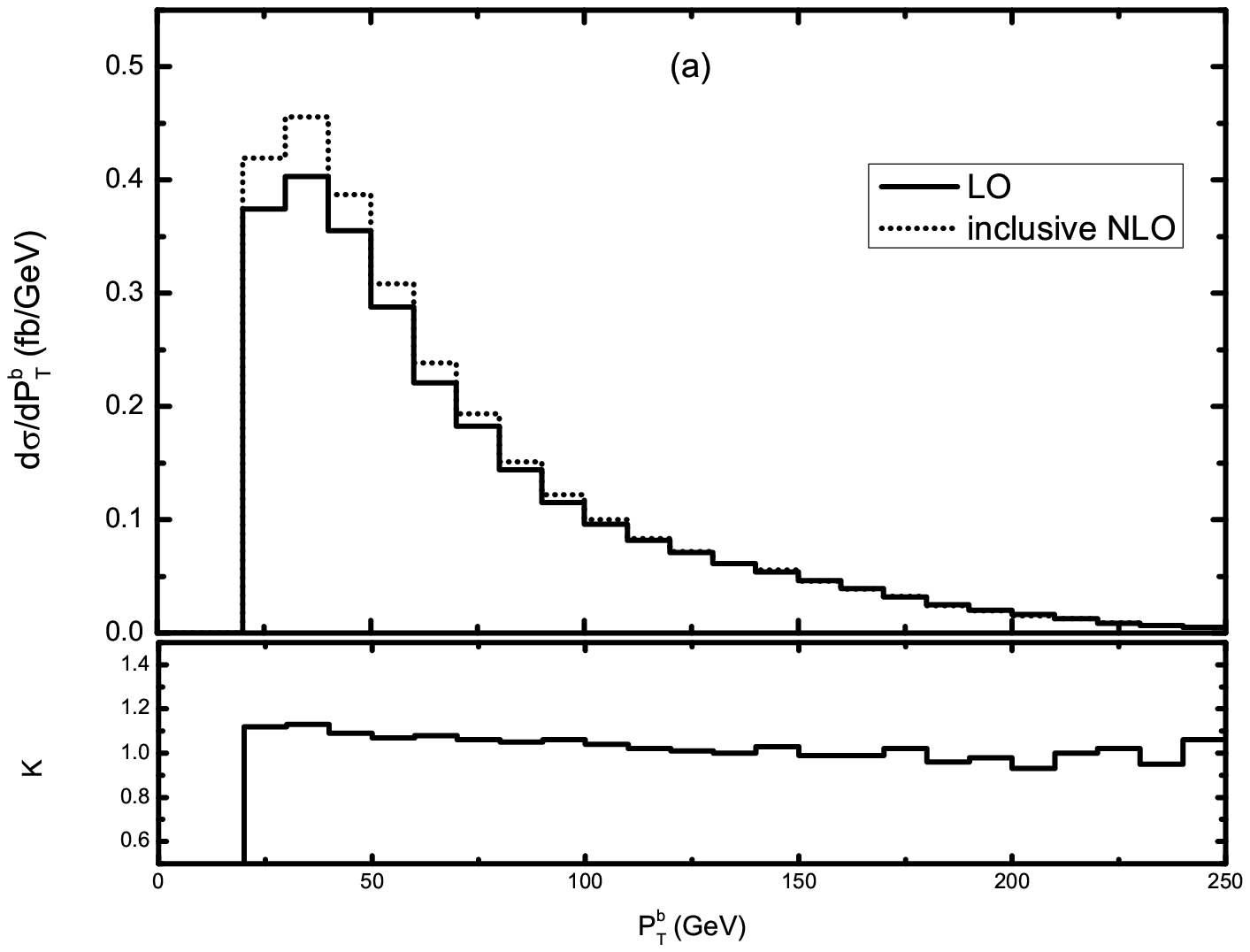}
\includegraphics[scale=0.53,bb=35 32 459 347]{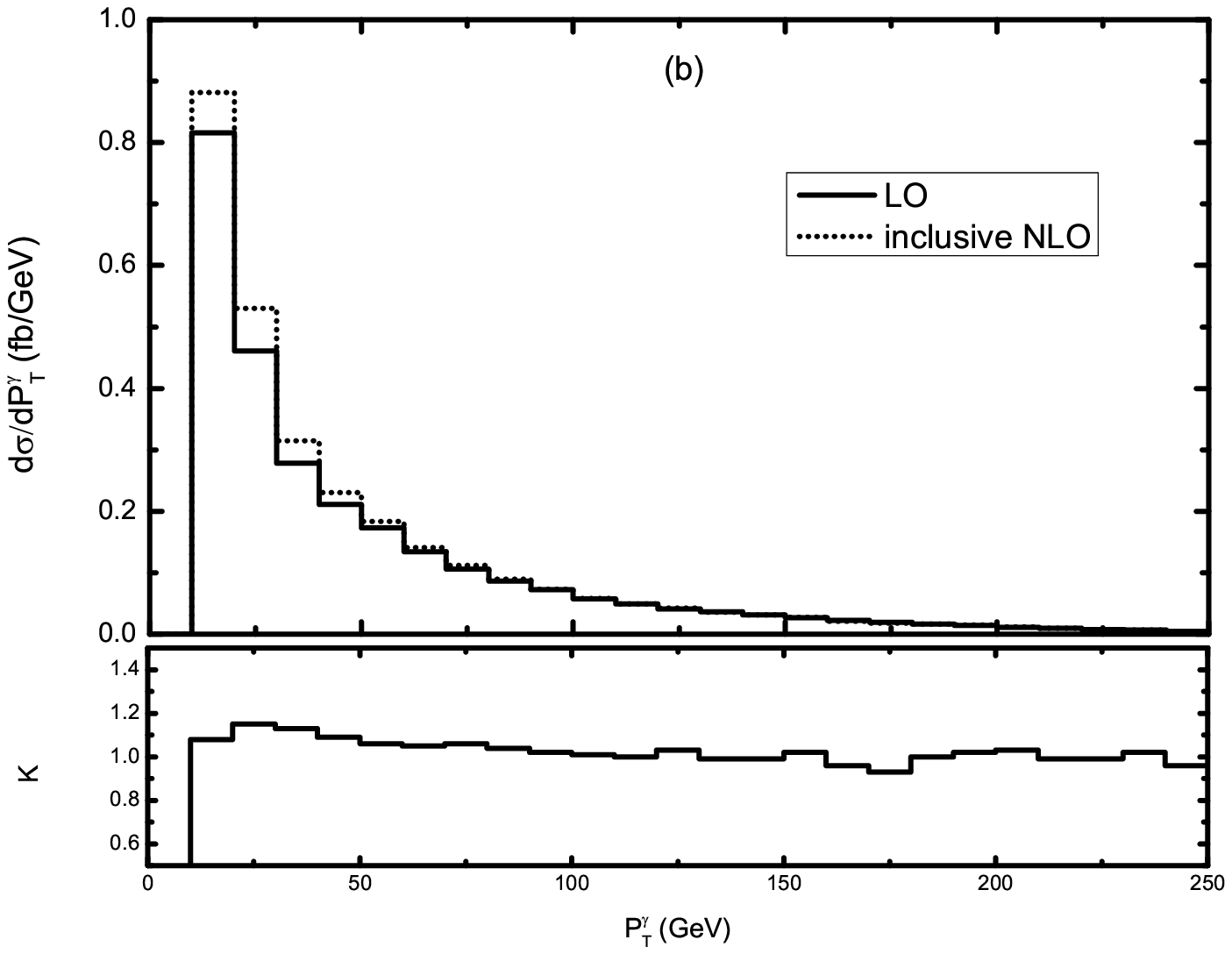}
\hspace{0in}%
\caption{\label{fig4} The LO and the ${\cal O}(\alpha_s\alpha^4)$
distributions of the transverse momenta of bottom-quark and
the leading photon, $p_{T}^{(b)}$, $p_{T}^{(\gamma)}$, in the
conditions of $\sqrt{s}=500~GeV$, $\mu=\sqrt{s}/2$ and the
'inclusive' selection scheme. There we take the cut values of
$p_{T,cut}^{(\gamma)}=10~GeV$, $\Delta R_{\gamma\gamma}^{cut}=0.5$,
$p_{T,cut}^{(b)}=20~GeV$ and $\Delta R_{b\bar b}^{cut}=\Delta
R_{b(\bar b)\gamma}^{cut}=1$. (a) The LO and ${\cal O}(\alpha_s)$ QCD corrected
distributions of transverse momentum of bottom-quark. (b) The LO and
${\cal O}(\alpha_s)$ QCD corrected distributions of transverse momentum of the
final leading photon. }
\end{center}
\end{figure}

Due to the CP-conservation, the $p_T^{(b)}$ distribution should be
the same as anti-bottom's ($p_T^{(\bar{b})}$). Here we present the
LO and QCD corrected distributions of the transverse momenta for the
bottom-quark and the leading photon with the 'inclusive' three-jet
event selection scheme in Fig.\ref{fig4}(a) and Fig.\ref{fig4}(b)
respectively, the corresponding K-factors are also plotted there.
The so-called leading photon is defined as the photon with the highest energy
among the two final photons. These results are obtained by taking
$\sqrt{s}=500~GeV$, $\mu=\sqrt{s}/2$ and the cut set for
b-quarks and photons as mentioned above. From these two figures we can see that the
${\cal O}(\alpha_s)$ QCD corrections enhance both the LO
differential cross sections $d\sigma_{LO}/dp_T^{(b)}$ and
$d\sigma_{LO}/dp_T^{(\gamma)}$, especially in low $p_T$ region. The
$p_T^{(\gamma)}$ distribution curves in Fig.\ref{fig4}(b) drop with
growing $p_T^{(\gamma)}$. Fig.\ref{fig4}(a) shows that the
differential cross sections ($d\sigma_{LO}/dp_T^{(b)}$,
$d\sigma_{NLO}/dp_T^{(b)}$) have their maximal values at about
$30~GeV \sim 40~GeV$, but Fig.\ref{fig4}(b) shows the maximal values
of $d\sigma_{LO}/dp_T^{(\gamma)}$ and $d\sigma_{NLO}/dp_T^{(\gamma)}$
are located at about $p_T^{(\gamma)}=10~GeV \sim 20~GeV$.

\par
\begin{figure}
\begin{center}
\includegraphics[scale=0.53,bb=35 32 459 347]{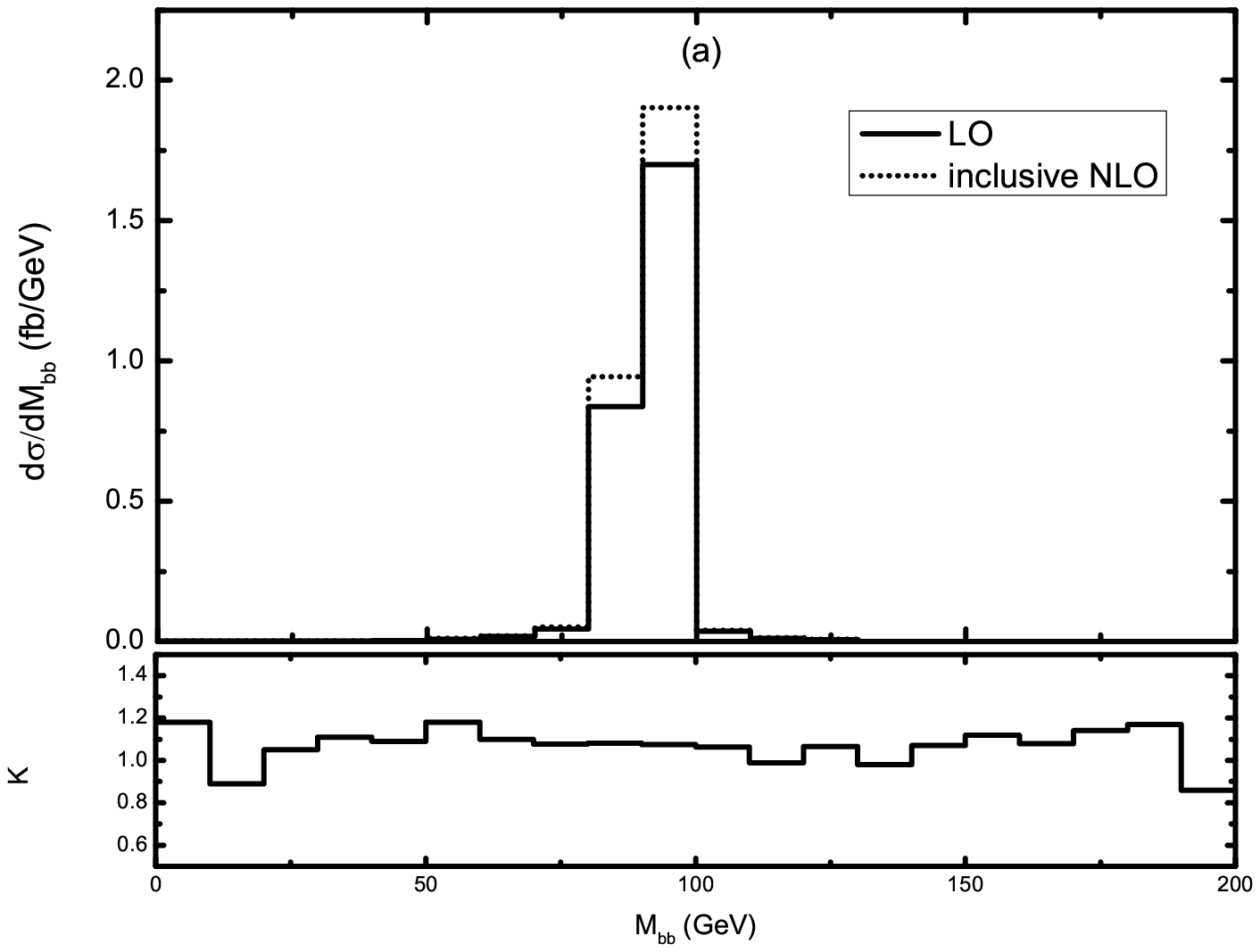}
\includegraphics[scale=0.53,bb=35 32 459 347]{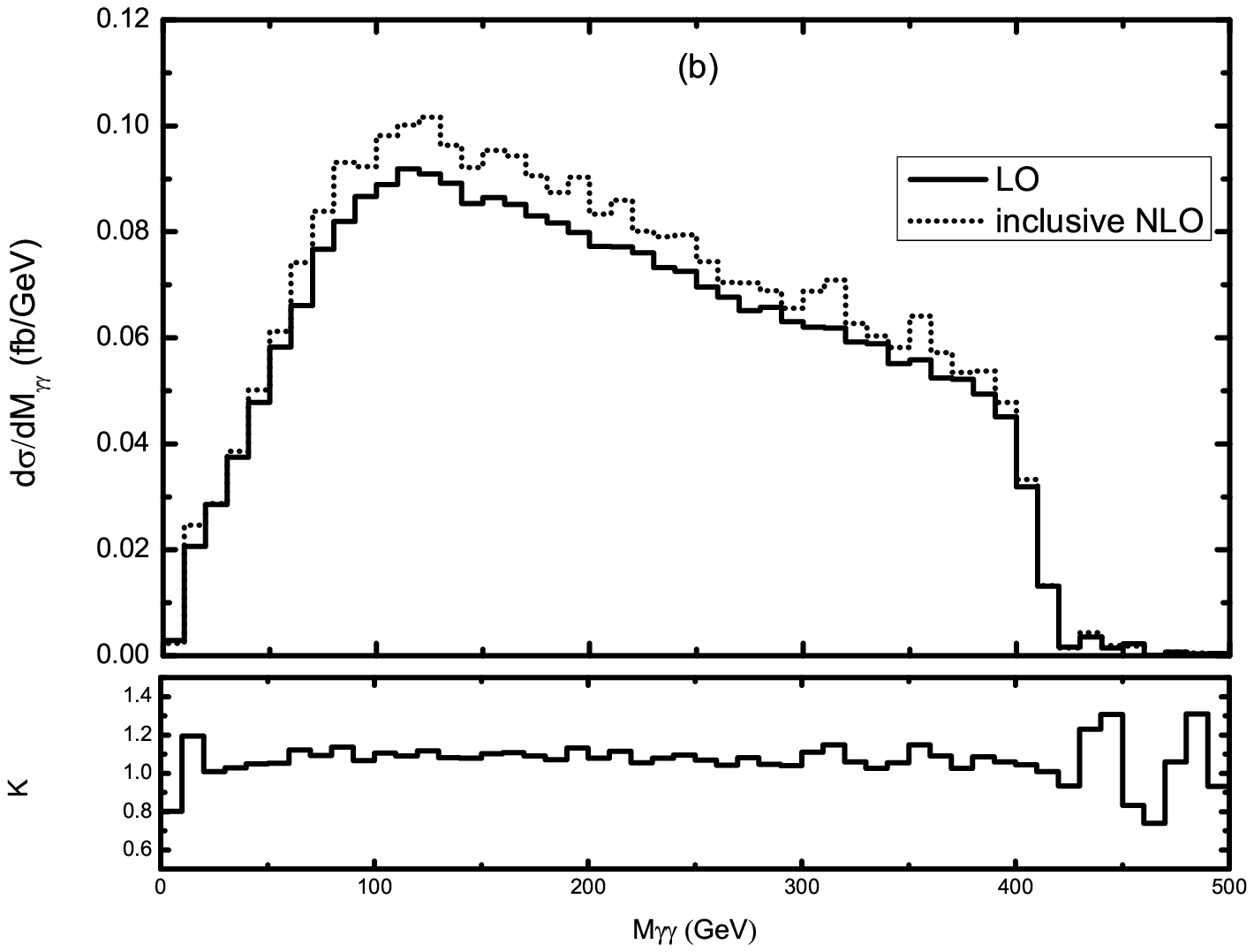}
\caption{\label{fig6} The distributions of the invariant masses of
$(b\bar b)$- and $(\gamma\gamma)$-pair at the LO and ${\cal
O}(\alpha_s\alpha^4)$ in conditions of $\sqrt{s}=500~GeV$,
$\mu=\sqrt{s}/2$, $p_{T,cut}^{(\gamma)}=10~GeV$, $\Delta
R_{\gamma\gamma}^{cut}=0.5$, $p_{T,cut}^{(b)}=20~GeV$ and $\Delta
R_{b\bar b}^{cut}=\Delta R_{b(\bar b)\gamma}^{cut}=1$. (a) The
distribution of the invariant mass of $(b\bar b)$-pair. (b) The
distribution of the invariant mass of $(\gamma\gamma)$-pair.  }
\end{center}
\end{figure}

We plot the spectra of $(b\bar b)$- and $(\gamma\gamma)$-pair
invariant masses (denoted as $M_{(b\bar b)}$ and
$M_{(\gamma\gamma)}$) with the 'inclusive' three-jet event selection
scheme at the LO and ${\cal O}(\alpha_s\alpha^4)$ in
Figs.\ref{fig6}(a) and (b), respectively. There we take
$\sqrt{s}=500~GeV$, $\mu=\sqrt{s}/2$, $p_{T,cut}^{(\gamma)}=10~GeV$, $\Delta
R_{\gamma\gamma}^{cut}=0.5$, $p_{T,cut}^{(b)}=20~GeV$ and $\Delta
R_{b\bar b}^{cut}=\Delta R_{b(\bar b)\gamma}^{cut}=1$.
We can see from Fig.\ref{fig6}(a) that most of
the events are concentrated around a peak located at the vicinity of
$M_{(b\bar b)} \sim m_Z$. That shows the fact that the contribution
to the cross section for the \eebbgg process at the ILC, is mainly
from real $Z^0$-boson production channel $e^+e^- \to \gamma\gamma
Z^0$ and followed by the subsequent real $Z^0$ decay $Z^0 \to
b\bar b$. Both the Figs.\ref{fig6}(a) and (b) show that the QCD
corrections enhance the LO differential cross sections
$d\sigma_{LO}/dM_{(b\bar b)}$ and
$d\sigma_{LO}/dM_{(\gamma\gamma)}$. The precise prediction for the
distribution of the $(\gamma\gamma)$-pair invariant mass is very
significant, because it is the irreducible continuum background for
the Higgs-boson signature of $H^0 \to \gamma\gamma$ decay in the
$\gamma\gamma b\bar b$ production process.

\vskip 5mm
\section{Summary}
In this paper we calculate the complete ${\cal O}(\alpha_s)$ QCD
corrections to the \eebbgg process in the SM at the ILC. We study
the dependence of the LO and ${\cal O}(\alpha_s)$ QCD corrected
cross sections on the colliding energy $\sqrt{s}$, and investigate
the LO and ${\cal O}(\alpha_s)$ QCD corrected distributions of the
transverse momenta of final particles and the spectra of the
invariant masses of $(\gamma\gamma)$- and $(b\bar b)$-pair. The
precise spectrum for the invariant mass of $\gamma\gamma$-pair is
very important, since it is the irreducible background if the Higgs
boson is produced via $e^+e^-\to H^0Z^0 \to \gamma\gamma b\bar b$
channel. Our calculations show that the size of the ${\cal
O}(\alpha_s)$ QCD correction exhibits a obvious dependence on the
additional gluon-jet veto scheme. The numerical results show that
the QCD corrections with 'inclusive' scheme enhance the LO results
by about $9.2\%$ to $7.0\%$ when we take the cut of
$p_{T,cut}^{(\gamma)}=10~GeV$, $\Delta R_{\gamma\gamma}^{cut}=0.5$,
$p_{T,cut}^{(b)}=20~GeV$, $\Delta R_{b\bar b}^{cut}=\Delta R_{b(\bar
b)\gamma}^{cut}=1$ with the colliding energy running from $200~GeV$
to $800~GeV$.

\vskip 3mm
\par
\noindent{\large\bf Acknowledgments:} This work was supported in
part by the National Natural Science Foundation of China
(Contract Nos.10875112, 10675110, 11005101, 11075150), and the
Specialized Research Fund for the Doctoral Program of Higher Education
(SRFDP)(No.20093402110030).

\end{document}